\documentclass[twocolumn,showpacs,preprintnumbers,amsmath,amssymb]{revtex4}
\usepackage{graphicx}
\usepackage{textcomp}

\begin{document}
%\draft
\title{Discrete diffraction and shape-invariant beams in optical waveguide arrays}
  \normalsize
\author{Stefano Longhi %\footnote{Author's email address: longhi@fisi.polimi.it}
}
\address{Dipartimento di Fisica and Istituto di Fotonica e
Nanotecnologie del CNR, Politecnico di Milano, Piazza L. da Vinci
32, I-20133 Milano, Italy}

%\date{.}
%
\bigskip
\begin{abstract}
\noindent
 General properties of linear propagation of discretized light in
 homogeneous and curved waveguide arrays are comprehensively
 investigated and compared to those of paraxial diffraction in
 continuous media. In particular, general laws describing beam spreading,
 beam decay and discrete far-field patterns in homogeneous arrays are derived using the method of
 moments and the steepest descend method. In curved arrays, the
 method of moments is extended to describe evolution of global
 beam parameters. A family of beams which propagate in curved arrays maintaining their functional
shape -referred to as discrete Bessel beams- is also introduced.
Propagation of discrete Bessel beams in waveguide arrays is simply
described by the evolution of a complex $q$ parameter similar to
the complex $q$ parameter used for Gaussian  beams in continuous
lensguide media. A few applications of the $q$ parameter formalism
are discussed, including beam collimation and polygonal optical
Bloch oscillations.
\end{abstract}

% 42.82.Et Waveguides, couplers, and arrays
% 42.79.Gn Optical waveguides and couplers

\maketitle

\section{Introduction}

Linear and nonlinear propagation of 'discretized' light in arrays
of evanescently-coupled optical waveguides has received a great
and increasing interest in the past recent years (see, for
instance, \cite{Christodoulides03,Lederer08} and references
therein). As compared to diffraction or refraction in continuous
(non-structured) media, discrete diffraction and refraction in
waveguide arrays show rather uncommon effects which result from
the evanescent coupling among adjacent waveguides forming a
one-dimensional or a two-dimensional lattice. For instance, linear
propagation of light waves in homogeneous arrays may show
diffraction reversal and self-collimation effects
\cite{Eisenberg00,Pertsch02}, anomalous refraction
\cite{Pertsch02}, the discrete Talbot effect \cite{Iwanow05}, and
quasi-incoherent propagation \cite{SzameitAPL07} to name a few.
Remarkably, discrete diffraction can be tailored by properly
introducing inhomogeneities in the lattice or by varying its
topology. In particular, since the first proposals and
demonstrations of optical Bloch oscillations
\cite{Peschel98,Morandotti99,Lenz99} and 'diffraction management'
in zig-zag arrays \cite{Eisenberg00}, the use of waveguide arrays
with curved optical axis has been extensively investigated both
theoretically and experimentally, with the demonstration of
diffraction suppression via  Bloch oscillations
\cite{Christodoulides03,Peschel98,Morandotti99,Lenz99,Usievich04}
or dynamic localization \cite{Lenz03,Longhi05}, polychromatic
diffraction management \cite{Garanovich06}, astigmatic diffraction
control \cite{Garanovich07}, multicolor Talbot effect
\cite{Garanovich06}, and discrete soliton management
\cite{Musslimani01}. Linear and nonlinear light propagation at the
surface or at the interface of two waveguide lattices also
exhibits a variety of interesting properties which have been
investigated in several recent works (see, for instance,
\cite{Lederer08,surf1,surf2,surf3} and references therein). In
spite of such a great amount of works, some facets of discrete
diffraction, even in the simplest linear propagation regime, have
been overlooked. Though in the linear regime the impulse response
(Green function) of the array may be rather generally calculated
analytically -either in straight or curved geometries and in
presence or not of boundaries- and its knowledge is enough to
predict light evolution for any assigned initial excitation
condition (see, for instance, \cite{Longhi05,surf1}), some general
issues of discrete diffraction, which are well known for paraxial
propagation of beams in continuous media, have not been
comprehensively addressed, including: (i) a description of global
beam parameter evolution in a closed analytical form; (ii)
far-field discrete diffraction in homogeneous array (the analogue
of Fraunhofer diffraction in homogeneous continuous media); (iii)
the general scaling law of beam broadening and beam decay,
especially close to the self-collimation condition (also referred
to as sub-diffraction) which is commonplace  to the more general
class of photonic crystal structures (see, for instance,
\cite{Kosaka99}); (iv) the existence of shape-invariant
discretized beams, i.e. special families of field distributions
which -like Gaussian beams in continuous lensguide media- do
propagate in straight or curved waveguide arrays maintaining their
functional shape.  It is the aim of this work to shed some light
into such issues. In particular, it is shown rather generally
that: (i) the scaling law describing broadening of discretized
light in homogeneous arrays is the same as that of standard
paraxial diffraction theory of homogeneous continuous media (beam
size asymptotically grows linearly with propagation distance),
independently of the precise array dispersion curve and even along
self-collimation directions; (ii) in a homogenous array, the
discrete far-field pattern is not the (discrete) Fourier transform
of the near-field distribution, and the scaling law of beam decay
may depend on the observation angle; (iii) special field
distributions, which propagate in straight or curved waveguide
arrays maintaining their functional shape and referred to as
'discrete Bessel beams', can be introduced for simple
tight-binding waveguide models; (iv) a discrete Bessel beam is
defined by a complex $q$ parameter, analogous to the one used for
Gaussian beams in continuous lensguide media, and propagation of
the $q$ parameter along the array admits of a simple geometric
interpretation.\\
The paper is organized as follows. In Sec.II general properties of
discrete diffraction in homogeneous waveguide arrays are
presented, including the derivation of the general scaling laws of
beam broadening and beam decay, far-field discrete diffraction,
with a a note on self-collimation regimes. In Sec.III, some
general rules of beam propagation in curved waveguide arrays are
derived within the nearest-neighbor coupling approximation,
whereas in Sec.IV the family of shape-invariant discrete Bessel
beams is introduced, together with the complex $q$ parameter
formalism. Applications to beam collimation and polygonal optical
Bloch oscillations are also presented. Finally, in Sec.V the main
conclusions are outlined.

\section{Discrete diffraction in a homogeneous waveguide array}
\subsection{Continuous model of discrete diffraction}
The starting point of our analysis is provided by a rather
standard model describing linear propagation of monochromatic
light waves along the $z$ direction of a one-dimensional or
two-dimensional array of waveguides  in the single band and
tight-binding approximations. For instance, in a one-dimensional
array such conditions are satisfied when the tilt of beams and
waveguides at the input facet is less than the Bragg angle, so
that the lowest-order band of the array is excited and beam
propagation is primarily characterized by coupling between the
fundamental modes of the waveguides. For a two-dimensional array,
the relevant equations describing discrete diffraction in a single
band approximation read
\begin{equation}
i \dot c_{n,m}=-\sum_{l,r} \Delta_{n-l,m-r} c_{l,r}
\end{equation}
where $c_{n,m}(z)$ is the complex amplitude of the fundamental
waveguide mode at the lattice site
$\mathbf{r}_{n,m}=n\mathbf{a}+m\mathbf{b}$ identified by the
indices $(n,m)$, $\mathbf{a}$ and $\mathbf{b}$ are the lattice
vectors of the unit cell,
 the dot denotes the derivative with respect
to $z$, and $\Delta_{n,m}=\Delta_{m,n}^*$ are the coupling rates.
In order to derive a general rule of beam broadening due to
discrete diffraction, it is worth introducing a continuous field
envelope $\psi(x,y,z)$ satisfying the scalar Schr\"{o}dinger-like
equation
\begin{equation}
i \partial_z \psi(\mathbf{r},z)=H_0(\mathbf{p}) \psi
\mathbf{(r},z),
\end{equation}
where $\mathbf{r}=(x,y)$, $\mathbf{p}=-i \nabla_{\mathbf{r}}$,
\begin{equation}
H_0(\mathbf{p}) \equiv -\sum_{n,m} \Delta_{n,m} \exp \left( -i
\mathbf{r}_{n,m} \cdot \mathbf{p} \right),
\end{equation}
and $\mathbf{r}_{n,m}=n\mathbf{a}+m\mathbf{b}$. Taking into
account that $\exp(-i \mathbf{R} \cdot \mathbf{p}) \psi
(\mathbf{r},z)=\psi(\mathbf{r+R},z)$, it follows that the solution
$c_{n,m}(z)$ to the discrete equation (1) can be identified with
$\psi(\mathbf{r}=n\mathbf{a}+m\mathbf{b},z)$. The formulation of
the discrete light propagation problem [Eq.(1)] as a continuous
problem [Eq.(3)] is a well-established procedure in solid-state
physics \cite{Ziman} which enable the use of certain analytical
techniques, such as the method of moments, developed for the
continuous Schr\"{o}dinger equation or for the paraxial wave
equation (see, for instance, \cite{Krivoshlykov,Styer90}). In
addition, the continuous model includes, as a particular case, the
problem of paraxial diffraction in a homogeneous medium (e.g. in
the vacuum), which is attained by simply assuming for the
Hamiltonian $H_0(\mathbf{p})$, in place of Eq.(3), the
(normalized) parabolic form
\begin{equation}
H_0(\mathbf{p})=\frac{\mathbf{p}^2}{2}.
\end{equation}
The normalization conditions $\int d\mathbf{r}
|\mathbf{\psi}\mathbf{(r},z)|^2=1$ for Eq.(2), and
$\sum_{n,m}|c_{n,m}(z)|^2=1$ for the discrete problem (1), will be
assumed in the following analysis.
\subsection{General law for beam spreading: moment analysis}
Two global parameters describing beam propagation are the beam
center of mass $\langle \mathbf{r} \rangle = \langle x \rangle
\mathbf{u}_x+\langle y \rangle \mathbf{u}_y$ and the transverse
beam spot sizes $w_x(z)$ and $w_y(z)$ defined by
\begin{equation}
\langle \mathbf{r} \rangle = \int d\mathbf{r} \mathbf{r}
|\psi(\mathbf{r},z)|^2,
\end{equation}
\begin{eqnarray}
w_x(z) & = & \sqrt{\langle x^2 \rangle -\langle x \rangle^2} \\
w_y(z) & = & \sqrt{\langle y^2 \rangle -\langle y \rangle^2}
\end{eqnarray}
where
\begin{equation}
\langle x^2 \rangle(z)=\int d\mathbf{r} x^2 |\psi(\mathbf{r},z)|^2
\; , \; \langle y^2 \rangle(z)=\int d\mathbf{r} y^2
|\psi(\mathbf{r},z)|^2.
\end{equation}
Note that the above definitions hold for both continuous and
discrete diffraction models. In the latter case, assuming
$\psi(\mathbf{r},z)$ to be a piecewise constant function in each
cell of the lattice and taking $|\mathbf{a} \times \mathbf{b}|=1$
for the area of the unit cell, the integral over $\mathbf{r}$ may
be replaced by a double sum over the cell indices $n$ and $m$,
i.e. in the discrete model one has  $\int d\mathbf{r}
\rightarrow \sum_{m,n}$. \\
The evolution equations for $\mathbf{r}$ and $w_{x,y}$ can be
readily obtained in a closed form by writing a set of Eherenfest
equations for the expectation values of $\mathbf{\mathbf{r}}$,
$x^2$, $y^2$, and of commutator operators that arise in the
calculation. The expectation value $\langle A \rangle \equiv \int
d\mathbf{r} \psi^*(\mathbf{r},z) A(\mathbf{r},\mathbf{p})
\psi(\mathbf{r},z)$ of any operator $A(\mathbf{r},\mathbf{p})$
(not necessarily self-adjoint) evolves according to
\begin{equation}
\frac{d \langle A \rangle}{dz}=-i \langle [A,H_0] \rangle
\end{equation}
and the following commutation relations
\begin{equation}
[\mathbf{r},f(\mathbf{p})]=i \nabla_{\mathbf{p}}f \; , \;
[g(\mathbf{r}),\mathbf{p}]=i \nabla_{\mathbf{r}}g
\end{equation}
hold for any functions $f(\mathbf{p})$ and $g(\mathbf{r})$. For
$A=\mathbf{r}$, one obtains
\begin{equation}
\frac{d \langle \mathbf{r}\rangle}{dz}=\langle \nabla_{\mathbf{p}}
H_0 \rangle , \; \frac{d \langle \nabla_{\mathbf{p}} H_0 \rangle
}{dz} =0,
\end{equation}
i.e.
\begin{equation}
\langle \mathbf{r} \rangle (z)=\langle \mathbf{r} \rangle
(0)+\langle \nabla_{\mathbf{p}} H_0 \rangle z
\end{equation}
which is the evolution equation of the beam center of mass. Note
that the path followed by any beam is always straight, regardless
of the specific form of $H_0$ or initial field distribution which
just determine the transverse drift velocity $\langle
\nabla_{\mathbf{p}} H_0 \rangle$ of the beam. To determine the
evolution equation of the beam spot size $w_x$, let us assume
$A=x^2$, so that the following cascade of Eherenfest equations
[Eq.(9)] is obtained
\begin{eqnarray}
\frac{d \langle x^2 \rangle}{dz} & = & \langle x \frac{\partial
H_0}{\partial p_x}+ \frac{\partial H_0}{\partial p_x} x \rangle \\
\frac{d}{dz} \langle x \frac{\partial H_0}{\partial p_x}+
\frac{\partial H_0}{\partial p_x} x  \rangle & = &  2 \langle
 \left( \frac{\partial H_0}{\partial p_x}\right)^2\rangle \\
 \frac{d}{dz}\langle \left( \frac{\partial H_0}{\partial
 p_x}\right)^2\rangle & = & 0.
\end{eqnarray}
After integration, one obtains
\begin{equation}
\langle x^2 \rangle(z) =\langle x^2 \rangle(0)+ \langle
  x \frac{\partial H_0}{\partial p_x}+
\frac{\partial H_0}{\partial p_x} x  \rangle z+  \langle \left(
\frac{\partial H_0}{\partial p_x}\right)^2 \rangle z^2,
\end{equation}
where the expectation values on the right hand side of Eq.(16) are
calculated at $z=0$, i.e. for the initial beam distribution. From
Eqs.(6), (12) and (16) the following evolution equation for the
beam spot size $w_x$ is then obtained
\begin{equation}
w_{x}^2(z)=w_{x}^2(0)+ \alpha_x z +\beta_x^2 z^2,
\end{equation}
where we have set
\begin{equation}
\alpha_x=\langle (x-\langle x \rangle ) \frac{\partial
H_0}{\partial p_x }+\frac{\partial H_0}{\partial p_x} (x-\langle x
\rangle ) \rangle,
\end{equation}
\begin{equation}
\beta_x^2=\langle \left( \frac{\partial H_0}{\partial
p_x}\right)^2\rangle-\left( \langle \frac{\partial H_0}{\partial
p_x}\rangle \right)^2
\end{equation}
and the expectation values are calculated at $z=0$. A similar
expression for $w_y(z)$ can be obtained by replacing $x$ with $y$
in Eqs.(17), (18) and (19). A major result expressed by Eq.(17) is
that, regardless of the particular form of $H_0$, $w_x(z)$ (and
similarly $w_y(z)$) asymptotically  grows with $z$ linearly, with
a diffraction length given by $\sim 1/\beta_x$. Therefore -and
this one of the major result of this section- the broadening law
of a spatial beam due to diffraction does not differ for discrete
or continuous diffraction. In addition, for a beam carrying a
finite power and admitting of finite moments $\langle x^2 \rangle$
and $\langle y^2 \rangle$, the coefficient $\beta_x^2$ given by
Eq.(19) is always strictly positive and does not vanish. This can
be generally proven by observing that $\beta_x^2$ is the variance
of the operator $(\partial H_0 /
\partial p_x)$, which is always positive and vanishes solely when
the initial field distribution $\psi(x,y,0)$ is an eigenfunction
of $(\partial H_0 /
\partial p_x)$, i.e. of $p_x=-i \partial_x$. Since such
eigenfunctions are delocalized plane waves, it then follows that
the variance of $(\partial H_0 / \partial p_x)$ is strictly
positive for any initial beam distribution carrying a finite
power, regardless of the specific form of $H_0$.

\subsection{Self-collimation regime}
Beam self-collimation (also referred to as beam sub-diffraction)
 is a well known phenomenon occuring in homogeneous arrays
and, more generally, in photonic crystal structures with
engineered band structure $H_0(\mathbf{p})$ showing points of
local flatness (see, for instance, \cite{Kosaka99}). The simplest
example of sub diffraction is the 'arrest' of beam spreading in a
one-dimensional tight-binding lattice with nearest-neighboring
couplings, which was observed in Ref.\cite{Pertsch02} using
relatively broad Gaussian beams  at an incidence angle set in
correspondence of an inflexion point of the band dispersion curve.
Though it is well understood that in such a regime diffraction is
cancelled solely at low orders, it was perhaps overlooked the fact
that self-collimation {\it does not} modify the beam broadening
scaling law [Eq.(17)]. In other words, self-collimation will
correspond to a reduction of the coefficient $\beta_x^2$ for
special initial field distributions, but {\it not} to a change of
the scaling law describing beam broadening. If we consider, for
the sake of simplicity, a one-dimensional lattice and assume that
the spectrum $F(k)$ of the exciting beam, defined as $F(k)=1/(2
\pi) \int dx \psi (x,0)\exp(-ikx)$, is narrow at around its mean
$k_0$, the value of $\beta_x^2$, as given by Eq.(19), can be
expanded in series of moments $I_l=\int dk (k-k_0)^l |F(k)|^2$
($l=2,3,4,...$) as
\begin{equation}
\beta_x^2=b_2^2 I_2+b_2b_3 I_3+\frac{1}{12} \left(4b_2b_4+3b_3^2
\right) I_4-\frac{b_3^2}{4} I_2^2+...
\end{equation}
where $b_l$ is the value of the derivative $(\partial ^l
H_0/\partial k^l)$ evaluated at $k=k_0$. As $I_l$ rapidly goes to
zero as the order $l$ increases, Eq.(20) shows that at the points
$k_0$ of the dispersion curve where $b_2$ (and possibly $b_3$,
$b_4$, ...) vanishes beam broadening is reduced. We will refer to
such points, where the dispersion curve $H_0(k)$ is locally flat,
to as self-collimation points [note that the condition $H'_0(k_0)=0$
is not requested].\\
 As an example, let us consider
the simplest one-dimensional waveguide array in the
nearest-neighboring approximation, considered in
Ref.\cite{Pertsch02} to demonstrate self-collimation effects. The
Hamiltonian $H_0$ has the form $H_0=-2 \Delta \cos (p)$, and the
self-collimation points are located at $p=\pm \pi/2$.  From
Eq.(19) one obtains
\begin{eqnarray}
\beta_x^2 & = & 2 \Delta^2 \left[1 - {\rm Re} \left(\sum_n c_n^*
c_{n+2} \right) \right] + \nonumber \\
& + & \Delta^2 \left[\sum_n c_n^* \left(c_{n+1}-c_{n-1} \right)
\right]^2.
\end{eqnarray}
For a bell-shaped (e.g. Gaussian-shaped) and flat beam incident
onto the array at a given tilting angle $\theta$ (normalized to
the Bragg angle), we may write $c_n= |c_n| \exp(-i \pi \theta n)$,
and one obtains
\begin{equation}
\beta_x^2(\theta)=2 \Delta^2 \left[
1-\kappa_1^2+(\kappa_1^2-\kappa_2) \cos(2 \pi \theta)\right]
\end{equation}
where $\kappa_1$ and $\kappa_2$ are defined by
\begin{equation}
\kappa_1 =  \sum_n|c_{n} c_{n+1}| \; , \; \kappa_2 =  \sum_n|c_{n}
c_{n+2}|.
\end{equation}
Generally, it turns out that $\kappa_1^2> \kappa_2$, so that the
minimum of $\beta_x$ is attained at $\theta= \pm 1/2$, i.e. at the
self-collimation points as expected from Eq.(20). Conversely, the
maximal diffraction (maximum value of $\beta_x$) is attained at
normal incidence ( $\theta=0$). The ratio between the minimum and
maximum values of $\beta_x$, given by
\begin{equation}
\Gamma=\frac{\beta_x(\theta=1/2)}{\beta_x(\theta=0)}=\sqrt{\frac{
1+\kappa_2 -2 \kappa_1^2 }{1-\kappa_2}},
\end{equation}
may get very small for a broad input beam. To illustrate this
point, let us consider as an example the following beam
distribution at the input plane : $|c_n|= \mathcal{N}
\alpha^{|n|}$, where the parameter $\alpha$ ($0<\alpha<1$)
determines the spot size of the input beam ($\alpha \rightarrow 0$
for single waveguide excitation, and $\alpha \rightarrow 1$ for a
plane wave excitation), and
$\mathcal{N}=[(1-\alpha^2)/(1+\alpha^2)]^{1/2}$ is a normalization
factor. For such a field distribution, the values of coefficients
$\kappa_1$ and $\kappa_2$ can be evaluated in a closed form, and
read
\begin{equation}
\kappa_1=\frac{2 \alpha}{1+\alpha^2} \; , \;
\kappa_2=\frac{\alpha^2(3- \alpha^2)}{1+\alpha^2}.
\end{equation}
The ratio $\Gamma$ between the diffraction parameters at
subdiffractive ($\theta=1/2$) and normal incidence ($\theta=0$)
regimes takes then the form [see Eq.(24)]
$\Gamma=[(1-\alpha^2)/(1+\alpha^2)]^{1/2}$. Note that, for a very
broad beam excitation ($\alpha \rightarrow 1$), both $\kappa_1$
and $\kappa_2$ gets close to 1, $\beta_x$ tends to vanish [see
Eq.(22)], and the diffraction length $ \sim 1 / \beta_x$ diverges
independently of beam tilting angle $\theta$, as expected for a
very broad input beam. However, in this case the ratio of
diffraction lengths in the normal ($\theta=0$) and subdiffractive
($\theta=1/2$) regimes, which scales as $\sim \Gamma$, tends to
vanish as $\Gamma \sim (1-\alpha)^{1/2}$. Conversely, for a very
narrow input beam ($\alpha \rightarrow 0$), both $\kappa_1$ and
$\kappa_2$ vanish and the diffraction length $\sim 1 / \beta_x$
turns out to be independent of tilting angle and given by $\sim
1/(\sqrt 2 \Delta)$ [see Eq.(22)] as expected for single waveguide
excitation.

\subsection{Discrete far-field pattern and anomalous beam decay}
In spite of the fact that the asymptotic law describing beam
broadening due to diffraction is the same for discrete and
continuous media, a deep difference is found when analyzing the
decay behavior of the field intensity versus propagation distance
and the far-field diffraction patterns.
 For the sake of simplicity, we will consider
the diffraction problem in one dimension, though the results may
be generalized to the two-dimensional diffraction problem. We then
rewrite Eq.(2) as
\begin{equation}
i \partial_z \psi(x,z)=H_0(p) \psi(x,z),
\end{equation}
where $p=p_x=-i \partial_x$. For the usual paraxial
one-dimensional diffraction problem in a homogeneous continuous
medium, one has $H_0(p)=p^2/2$, whereas for discrete diffraction
in a one-dimensional waveguide array one has $H_0(-p)=H_0(p)$
($-\pi \leq p <\pi$) and $H'_0(p)=0$ at $p=0$ and at the band
edges $p=\pm \pi$. The most general solution to Eq.(26) can be
written as
\begin{equation}
\psi(x,z)={\int} dk F(k) \exp[ikx-iH_0(k)z]
\end{equation}
 where the spectrum $F(k)$ is determined by the beam
 distribution at the input plane $\psi(x,0)$
 \begin{equation}
 F(k)=\frac{1}{2 \pi} \int dx \psi(x,0) \exp(-ikx)
 \end{equation}
($\int dx \rightarrow \sum_n$, $x \rightarrow n$ and $\psi(x=n)
\rightarrow c_n$ in the discrete diffraction problem). Our aim is
to calculate the decay behavior of the field amplitude $\psi(x,z)$
as the propagation distance $z$ increases, at either a constant
$x$ position (for instance at $x=0$) or along a line $x=\alpha z$,
where $\alpha$ is a constant parameter defining the 'observation
angle' of the diffracted pattern. Note that, as the observation
angle $\alpha$ is varied, the function $\psi_0(\alpha;z)=\psi
(x=\alpha z, z)$ corresponds, for large values of $z$, to the far
field diffraction pattern. We need thus to calculate the
asymptotic behavior of the integral
\begin{equation}
\psi_0(\alpha;z)={\int} dk F(k) \exp[iz g(k)]
\end{equation}
for $z \rightarrow \infty$, where we have set
\begin{equation}
g(k)=\alpha k-H_0(k).
\end{equation}
For this purpose, we may use the methods of stationary phase or
steepest descend (see, for instance, \cite{Oliver74}), which
predict that the asymptotic behavior of $\psi_0(\alpha;z)$ as $z
\rightarrow \infty$ depends on the existence and of the order of
stationary points of the phase $g(k)$ inside the
integration domain.\\
Let us first consider the continuous diffraction problem,
$H_0(p)=p^2/2$, and re-derive the well-known result that the
amplitude $\psi_0(\alpha;z)$ decays as $\sim 1 \sqrt{z}$ for any
observation angle $\alpha$ and the far-field pattern is
proportional to the Fourier spectrum of the input (near-field)
distribution. In this case, $g(k)=\alpha k-k^2/2$ has a unique
saddle point at $k=k_0=\alpha$, with $g''(k_0)=-1 \neq0$;
therefore, provided that the spectrum $F(k)$ has a nonvanishing
component at $k=k_0$ and $F(k_0)$ does not diverge \cite{note},
according to the method of steepest descend one has
\begin{equation}
\psi_0(\alpha; z) \sim F(\alpha)  \sqrt{\frac{2 \pi }{z}} \exp[iz
\alpha^2/2 -i \pi/4]
\end{equation}

\begin{figure}
\includegraphics[scale=0.6]{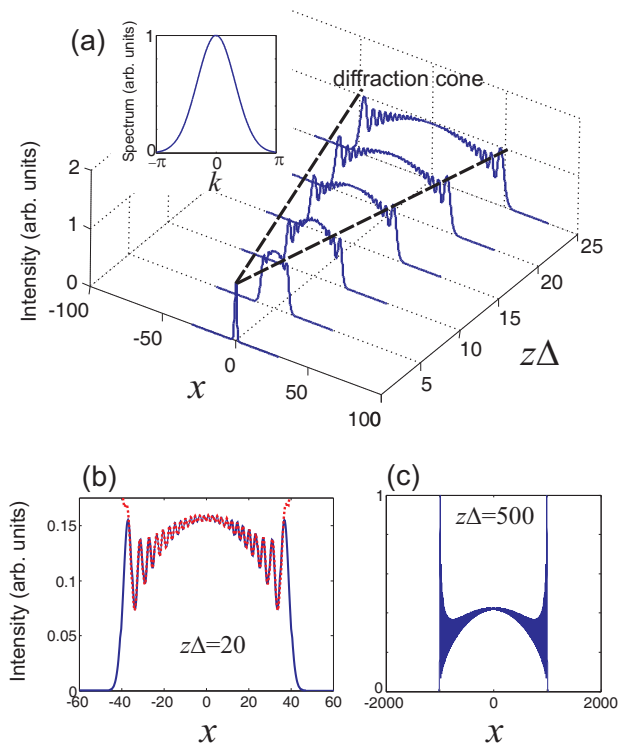}
\caption{(color online) Beam propagation in a one-dimensional
tight binding lattice
 with nearest neighboring coupling terms, showing far-field properties of discrete diffraction.
In (a) the intensity distributions $|\psi(x,z)|^2$ are plotted, in
arbitrary units, for propagation distances $z=0$, $z=5/ \Delta$,
$z=10 /\Delta$, $z=15/ \Delta$, $z=20/ \Delta$, and $z=25/
\Delta$, where $\Delta$ is the coupling rate between adjacent
waveguides. The inset in (a) shows the Gaussian spectrum $F(k)$ of
the beam ($w_k=1.5$). In (c) the intensity distribution
$|\psi(x,z)|^2$ is depicted for a propagation distance
$z=20/\Delta$ as numerically calculated by Eq.(27) (solid curve)
and by the approximate relation (33) (dotted curve, almost
overlapped with the solid one). In (c) the beam intensity is
plotted at a propagation distance $z=500/\Delta$, clearly showing
the dominance of two peaks at the diffraction cone edges
(self-collimation directions) and the onset of three different
decay laws at $|\alpha|>2 \Delta$, $\alpha=\pm 2 \Delta$, and
$|\alpha|<2 \Delta$.}
\end{figure}

as $z \rightarrow \infty$. From Eq.(31) we obtain the well-known
result of paraxial diffraction theory  that the amplitude
$\psi_0(\alpha;z)$ of the beam decays as $\sim 1 \sqrt{z}$ for any
observation angle $\alpha$ \cite{note}, and that the far-field
diffraction pattern is  shaped as the Fourier spectrum $F(\alpha)$
of the near-field distribution. This scaling law may be referred
to as the {\it normal} scaling law, in the sense that the beam
intensity $I \propto |\psi|^2$ decays as $\sim 1/z$ whereas the
beam spot size $w_x$ increases asymptotically as $\sim z$ [see
Eq.(17)], the product $I w_x$
being constant according to the power conservation law. \\
 For the discrete diffraction problem, we prove now that the decay law is generally
{\it slower} than $\sim 1 / \sqrt{z}$ at the observation angles
corresponding to self-collimation, and that the far-field pattern
is peaked at such angles and does not reproduce the spectrum $F$
of the near-field distribution. To this aim, let us observe that,
according to the steepest descend method, the slowest decay term
in the integral of Eq.(29) comes from the saddle points
$g'(k_0)=0$ of largest order. In particular, if $k_0$ is a saddle
point of order
 $n \geq 2$, i.e. $g(k) \simeq g(k_0)+[g^{(n)}(k_0)/n
!](k-k_0)^n$ for $k$ close to $k_0$ ($g^{(n)}(k_0) \neq 0$), the
contribution of the saddle point to the integral in Eq.(29) for
large values of $z$ is given by \cite{Oliver74}
\begin{equation}
\psi_0(\alpha;z) \sim  \frac{F(k_0)}{|zg^{(n)}(k_0)|^{1/n}} (n
!)^{\frac{1}{n}} \Gamma \left( \frac{1}{n} \right) \exp \left[ iz
g(k_0) \pm i \frac{\pi}{2n} \right].
\end{equation}
Therefore, the decay law for $\psi_0(\alpha;z)$ scales as $\sim
z^{-1/n}$, where $n$ is the highest order of the saddle points of
$g(k)$, provided that $F(k_0) \neq 0$. In the case of diffraction
in a homogeneous continuous medium, the order of saddle point is
always $n=2$. To determine $n$ for the discrete diffraction
problem, let us note that the dispersion curve $H_0(k)$ admits of
at least a couple of inflection points, say at $k=\pm k_0$, at
which $H''_0(k_0)=0$. These points correspond to the
self-collimation directions introduced in Sec.II.C. Since
$g'(k)=\alpha-H'_0(k)$, the inflection points $k= \pm k_0$ turn
out to be also saddle points when the observation angle $\alpha$
is chosen equal to $H'_0( \pm k_0)$. Therefore, for the discrete
diffraction problem the largest order $n$ of saddle points is {\it
at least} $n=3$, and the decay law of $\psi_0(\alpha;z)$, at the
two observation angles $\alpha= \pm H'_0(k_0)$ corresponding to
the self-collimation directions $ \pm k_0$, is slowed down -as
compared to continuous diffraction- to (at least) $\sim z^{-1/3}$.
More generally, if the dispersion curve $H_0(k)$ of the lattice is
engineered to achieve a very flat behavior near a self-collimation
point $k=k_0$, with $g''(k_0)=g'''(k_0)=...=g^{(n-1)}(k_0)=0$ and
$g^{(n)}(k_0) \neq 0$ ($n \geq 3$), the decay law of
$\psi_0(\alpha;z)$ scales as $\sim z^{-1/n}$ at the observation
angle $\alpha=H'(k_0)$.
 This scaling law of beam decaying
in the discrete diffraction problem is therefore {\it anomalous},
in the sense that along the self-collimation directions the
intensity decays slower that $1/z$, i.e. of the characteristic
decay law that one might expect from power conservation arguments.
This seemingly paradoxical circumstance may be solved by observing
that, for an observation angle $\alpha$ different than any of the
self-collimation directions, the decay of $\psi_0(\alpha;z)$ may
be either normal (i.e. $\sim 1/ \sqrt{z}$) or even {\it faster}.
More precisely, for a fixed value of $\alpha$ in modulus larger
than $\alpha_{max}={\rm max}_k |H_{0}^{'}(k)|$, the function
$g(k)$ given by Eq.(30) does not have saddle points on the real
axis, and $\psi_0(\alpha; z)$ decays as $ \sim 1/z$. Conversely,
for $|\alpha|< \alpha_{max}$ the equation
$g'(k)=\alpha-H_{0}^{'}(k)=0$ admits of at least one solution,
with $g''(k) \neq 0$ for a second-order saddle point. In this
case, according to the method of stationary phase the asymptotic
behavior of $\psi_0(\alpha;z)$ for large values of $z$ follows the
normal law $\sim 1 / \sqrt{z}$. To summarize, $\psi_0(\alpha;z)$
scales: as $\sim F(k_0) z^{-1/n}$ at a self collimation direction
$\alpha$, where $H_{0}^{'}(k_0)=\alpha$ and $k_0$ is a saddle
point of order $n \geq 3$; as $\sim F(k_0) z^{-1}$ for an
observation angle $\alpha$ outside the 'diffraction cone'
$|\alpha|>\alpha_{max}$;  as $\sim F(k_0) z^{-1/2}$ inside the
diffraction cone region $|\alpha|<\alpha_{max}$ but far from a
self collimation direction. The far-field pattern of discrete
diffraction tends therefore to confine light inside the
diffraction cone $|\alpha| \leq \alpha_{max}$ with asymptotic
peaks at the propagation directions corresponding to the angles of
self-collimation.\\
 This very general behavior may be illustrated
more in details for the case of a tight-binding lattice in the
nearest-neighbor approximation considered in Sec.II.C, for which
$H_0(k)=-2 \Delta \cos k$. In this lattice model one has
$g'(k)=\alpha-2\Delta \sin k$, $g''(k)=-2\Delta \cos k$, so that
the angle of diffraction cone is given by $\alpha_{max}=2\Delta$.
Two saddle points of second-order are found at $k=k_0= \pm \pi/2$
for the observation angles $\alpha=\pm \alpha_{max}$, i.e at the
edge of the diffraction cone, at which the far-field discrete
diffraction pattern is thus expected to show two peaks. For an
observation angle $\alpha$ {\it strictly} inside the diffraction
cone ($|\alpha|<2 \Delta$), the equation $g'(k)=0$ has two
solutions which are saddle points of first order since $g''(k)
\neq 0$. The main contribution to the integral on the right hand
side of Eq.(29) comes from these two saddle points, and can be
calculated by the method of stationary phase, yielding explicitly
\begin{widetext}
\begin{eqnarray}
\psi(\alpha;z)  & \sim &  \sqrt{ \frac{ i \pi}{z
\sqrt{\Delta^2-(\alpha/2)^2}}} \left\{-i F(k_0) \exp \left[i
\alpha k_0 z + 2i \Delta z \cos k_0\right] + F(\pi-k_0) \exp
\left[i \alpha z (\pi-k_0)- 2i \Delta z \cos k_0 \right]
\right\} \nonumber \\
& & (\alpha>0) \nonumber \\
\psi(\alpha;z)  & \sim &  \sqrt{ \frac{i \pi}{z
\sqrt{\Delta^2-(\alpha/2)^2}}} \left\{-i F(-k_0) \exp \left[-i
\alpha k_0 z + 2i \Delta z \cos k_0 \right] + F(-\pi+k_0) \exp
\left[i \alpha z (-\pi+k_0)- 2i \Delta z \cos k_0 \right]
\right\} \nonumber \\
& & (\alpha<0)
\end{eqnarray}
\end{widetext}
where $k_0$ is the solution to the equation $\sin k_0=|\alpha/ 2
\Delta|$ in the interval $ 0 \leq k_0 < \pi/2$. It should be noted
that the far-field discrete diffraction pattern given by Eq.(33)
holds for $|\alpha|<2 \Delta$. As $|\alpha|$ approaches $2 \Delta$
from below, the two saddle points of second order coalesce into a
single saddle point of third order, and this explain the
divergence of Eq.(33) as $|\alpha| \rightarrow 2 \Delta$, i.e at
the self collimation directions, where the decay is slower and
scales as $\sim z ^{-1/3}$. For $|\alpha|> 2 \Delta$, i.e. outside
the diffraction cone, there are not saddle points on the real axis
and the decay is faster and scales as $\sim 1/z$. An example of
far-field discrete diffraction for a beam with a Gaussian spectrum
$F(k) \sim \exp[-(k/w_k)^2]$ ($-\pi \leq k <\pi$) is shown in
Fig.1. In Fig.1(a) the intensity distribution $|\psi(x,z)|^2$, as
obtained by  accurate numerical computation of the integral
entering in Eq.(27), is plotted for a few propagation distances
$z$. For the sake of readability, at each propagation distance $z$
the field intensity has been normalized to its peak value. The
diffraction cone and the emergence of two intensity peaks at the
self-collimation directions $\alpha= \pm \alpha_{max}$ are clearly
visible just after a propagation distance $z$ of $\sim 10-20$
times the coupling length $1/\Delta$ [Fig.1(a)]. Inside the
diffraction cone, the intensity distribution at such propagation
distances is very well fitted by the analytical far-field
distribution given by Eq.(33), as shown in Fig.1(b). At much
longer propagation distances, the self-collimation peaks become
dominant, and the appearance of three different scaling laws of
beam decay (fast decay outside the diffraction cone $|x|>2 \Delta
z$; normal decay inside the diffraction cone $|x|<2 \Delta z$;
slower decay at the self-collimation directions $x= \pm 2 \Delta
z$) is very clearly visible, as shown in Fig.1(c).

\section{Beam propagation in curved waveguide arrays}

Discrete diffraction of light waves in linear optical waveguide
arrays can be controlled by introducing transverse index gradients
or local phase slips, which may produce a kind of refocusing or
re-imaging of beam distributions along the propagation distances
(see, for instance,
\cite{Lenz99,Eisenberg00,Lenz03,Longhi05,SzameitAPL08}) similarly
to what happens to light propagating in continuous lensguide
media. In particular, waveguide arrays with a curved axis provide
a particularly interesting set up to manage discrete diffraction
for both monochromatic and polychromatic light
\cite{Lenz99,Lenz03,Longhi05,Garanovich06,Garanovich07}. It is
therefore of major interest to have general laws describing the
global behavior of beam propagation in curved waveguide arrays. In
addition, it is well known that for the problem of paraxial
diffraction in homogeneous continuous media or, more generally, of
paraxial propagation in elementary optical systems and lensguides,
one can introduce special families of field distributions (such as
the Gaussian beams) that propagate maintaining unchanged their
functional shape (shape-invariant beams), and that field
propagation  may be simply described by means of algebraic
equations ruling out the evolution of some complex-valued beam
parameters (such as the complex $q$-parameter for Gaussian beams;
see, for instance, \cite{Siegman}). A natural question is whether
one can similarly introduce shape-invariant discrete beams, i.e.
field distributions that do not change their functional shape when
propagating along curved waveguide arrays. As the problem of
discrete diffraction in waveguide arrays with curved axis or
transverselly-imposed index gradients is analogous to the problem
of one-dimensional or two-dimensional Bloch oscillations of
electrons in periodic potentials with an applied electric field or
of cold atoms in optical lattices, some results are already
available in the literature. In particular, in recent works
\cite{B01,B02,B03} an algebraic approach has been developed,
capable of providing rather general results for wave packet center
of mass evolution and wave packet spreading in certain lattice
models. In this approach, after the introduction of a dynamical
Lie algebra, an explicit form of the evolution operator is first
derived, and then the expectation values of operators are
calculated in the Heisenberg picture. However, the question of
existence of shape-invariant discrete beams and of their
propagation in curved waveguide arrays does not seem to have been
addressed yet. In this section, we present a generalization of
Eqs.(12) and (17) describing the evolution of beam center of mass
and beam width in curved waveguide arrays using the method of
moments. Though similar results have been previously published in
Refs.\cite{B01,B02,B03} using an algebraic operator approach, they
are here re-derived for the sake of completeness using the method
of moments, which does not require the explicit calculation of the
evolution operator and the formulation of the problem in terms of
a Lie algebra. In the subsequent  section a family of
shape-invariant discrete beams will be introduced, proving that
their propagation in a generally-curved waveguide array is simply
described by the evolution of a complex-$q$ beam parameter, which
plays an analogous role of e.g. the complex-$q$ parameter of
Gaussian beams
propagating in paraxial continuous optical systems.\\
Let us consider monochromatic light propagation in a
two-dimensional waveguide array with a curved axis described by
the parametric equations $x=x_0(z)$ and $y=y_0(z)$; the coupled
mode equations describing light transfer among coupled waveguides
in the single-band and tight-binding approximations are an
extension of Eq.(1) to include fictitious transverse index
gradients induced by waveguide curvature and read explicitly
\begin{equation}
i \dot c_{n,m}=-\sum_{l,r} \Delta_{n-l,m-r}
c_{l,r}-\mathcal{E}(z)\cdot \mathbf{r}_{n,m}
\end{equation}
where
$\mathcal{E}(t)=\mathcal{E}_x(t)\mathbf{u}_x+\mathcal{E}_y(t)\mathbf{u}_y$,
$\mathcal{E}_x(z)=-(n_s/ \lambdabar) \ddot{x}_0(z)$,
$\mathcal{E}_y(z)=-(n_s/ \lambdabar) \ddot{y}_0(z)$, $n_s$ is the
refractive index of the waveguide substrate, and $\lambdabar=
\lambda/(2 \pi)$ is the reduced wavelength of light. It should be
noticed that the transverse index gradient entering in Eq.(34) may
be also realized by applying a thermal gradient, or may describe
lumped phase gradients \cite{SzameitAPL08} or an abrupt tilt of
waveguide axis direction \cite{Eisenberg00}, in which cases
$\mathcal{E}(z)$ shows a delta-like behavior. After introduction
of a continuous function $\psi(\mathbf{r},z)$ such that
$\psi(\mathbf{r}_{n,m},z)=c_{n,m}(z)$, one can readily check that
the discrete diffraction equations (34) are equivalent to the
following continuous Hamiltonian problem
\begin{equation}
i \partial_z \psi (\mathbf{r},z)=H(\mathbf{r},\mathbf{p}) \psi
(\mathbf{r},z)
\end{equation}
($\mathbf{p}=-i\nabla_{\mathbf{r}}$) with Hamiltonian
\begin{equation}
H=H_0(\mathbf{p})-\mathcal{E}(z) \cdot \mathbf{r},
\end{equation}
where $H_0$ is the Hamiltonian of the homogeneous array defined by
Eq.(3). The laws governing the evolution of beam center of mass
and beam variance can be obtained by extending the method of
moments described in Sec.II.B for the free diffraction problem. In
general, the cascade of equations that one obtains by applying the
Eherenfest equation (9) to $\langle \mathbf{r}\rangle$, $\langle
x^2 \rangle$, $\langle y^2 \rangle$ - and to the commutators found
throughout the calculations- turns out to be unlimited for a
general form of $H_0$, and a closed set of equations are found
solely for special forms of $H_0$. Such a special circumstance is
encountered in case of a one-dimensional waveguide array in the
nearest-neighboring approximation, and in case of a
rectangular-lattice waveguide array neglecting diagonal
interactions. The first model corresponds to the Hamiltonian
\begin{equation}
H(x,p)=-2 \Delta \cos p-\mathcal{E}_x(z)x,
\end{equation}
where $\Delta$ is the coupling rate between adjacent waveguides,
and $p=p_x=-i\partial_x$. The second model, which has been for
instance considered in the experiment of Ref.\cite{Pertsch04}, is
described by the Hamiltonian
\begin{equation}
H(\mathbf{r},\mathbf{p})=-2 \Delta_x \cos (p_x)-2\Delta_y
\cos(p_y) - \mathcal{E}_x(t)x-\mathcal{E}_y(z)y,
\end{equation}
where $\Delta_x$ ($\Delta_y$) is the coupling rate between
adjacent horizontal (vertical) waveguides of the lattice
\cite{notaS}.

\subsection{One-dimensional array}
Application of the moment method to the one-dimensional
Hamiltonian model (37) yields a set of closed coupled equations
for the expectation values of operators $x$, $\theta$, and of
$x^2$, $\rho$ and $\sigma$, where
\begin{eqnarray}
\theta & = & \exp(ip)  \\
\rho & = & \frac{1}{2} \left\{ 1-\exp(-2ip) \right\} \\
\sigma & = & i \left\{ x \exp(-ip)+\exp(-ip) x \right\}.
\end{eqnarray}
Successive application of the Ehrenfest equation (9) yields the
following equations for $\langle x \rangle$ and $\langle \theta
\rangle$
\begin{eqnarray}
\frac{d\langle x \rangle}{dz} & = & 2 \Delta {\rm Im}
\left(\langle \theta \rangle  \right) \\
\frac{d\langle \theta \rangle}{dz} & = & i \mathcal{E}_x(z)
\langle \theta \rangle  ,
\end{eqnarray}
and the following coupled equations for $\langle x^2 \rangle$,
$\langle \rho \rangle$ and $\langle \sigma \rangle$
\begin{eqnarray}
\frac{d\langle x^2 \rangle}{dz} & = & 2 \Delta {\rm Re}
\left(\langle \sigma \rangle  \right) \\
\frac{d\langle \rho \rangle}{dz} & = & -2 i \mathcal{E}_x(z)
\langle \rho \rangle +i \mathcal{E}_x(z) \\
\frac{d\langle \sigma \rangle}{dz} & = & 4 \Delta \langle \rho
\rangle -i \mathcal{E}_x(z) \langle \sigma \rangle.
\end{eqnarray}
Equation (43) can be readily integrated, yielding the following
evolution equation for the beam center of mass
\begin{equation}
\langle x(z) \rangle =\langle x(0) \rangle+2 {\rm Im} \left\{ q_0
\Omega^*(z) \right\}
\end{equation}
where we have set
\begin{eqnarray}
\Omega(z)  & \equiv  & \int_0^z d\xi \Delta \exp [-i \phi(\xi)],
\\
\phi(z) & \equiv &  \int_0^z d \xi \mathcal{E}_x(\xi) , \\
q_0 & \equiv  & \sum_n c_n^*(0)c_{n+1}(0).
\end{eqnarray}
Similarly, integration of Eqs.(45) and (46) yields
\begin{eqnarray}
\langle \rho(z) \rangle =   \exp[-2i \phi(z)] \times \nonumber
\\
 \times  \left\{ \langle \rho (0) \rangle +\frac{1}{2} \exp[2i
\phi(z)]-\frac{1}{2}\right\}
\\\
\langle \sigma(z) \rangle  =    \exp[-i \phi(z)]  \times
\nonumber \\
 \times   \left\{ \langle \sigma(0) \rangle +4 \Omega(z)
\left[\langle \rho(0) \rangle -\frac{1}{2} \right]+2 \Omega^*(z)
\right\}
\end{eqnarray}
Taking into account that $\Delta \exp[-i\phi(z)]=d \Omega / dz$
and that $2 {\rm Re} \left\{  \int_0^z d \xi  \Omega^*(\xi) (d
\Omega / d \xi)\right\}=|\Omega(z)|^2 $, substitution of Eq.(52)
into Eq.(44) yields
\begin{equation}
\langle x^2 (z)\rangle=\langle x^2 (0) \rangle+2|\Omega(z)|^2+2
{\rm Re} \left\{ q_1 \Omega(z) - q_2 \Omega^2(z) \right\},
\end{equation}
where we have set
\begin{eqnarray}
q_1 & \equiv & i \sum_n
(2n-1)c^*_n(0)c_{n-1}(0) \\
q_2 & \equiv & \sum_{n}c^*_n(0) c_{n-2}(0).
\end{eqnarray}
The beam size $w_x$ is then given by
\begin{equation}
w_x(z)= \sqrt{\langle x^2(z) \rangle -\langle x(z) \rangle^2}.
\end{equation}
 For a given field
distribution $c_n(0)$ at the input plane, the evolution of the
beam center of mass $\langle x (z) \rangle$ and beam size $w_x(z)$
are thus ruled by Eqs.(47), (53) and (56). Note that beam
evolution depend on the input beam parameters $q_1$, $q_2$ and
$q_3$ -defined by Eqs.(50),(54) and (55)- and by the complex
amplitude $\Omega(z)$, defined by Eqs.(48-49) and accounting for
bending of waveguide axis. Note also that for straight arrays
$\Omega(z)=\Delta z $, and one thus retrieves the results of
discrete diffraction derived in Sec.II.B, in particular the linear
asymptotic increase of $w_x$ with $z$. The condition for
diffraction suppression, i.e. a non-secular growth of $w_x(z)$
with $z$, is that $\Omega(z)$ remains a limited function of $z$ as
$z$ increases. This condition is always satisfied for a constant
value of  $\mathcal{E}_x$, which corresponds to circularly-curved
waveguides and to the onset of the optical analogue of Bloch
oscillations \cite{Lenz99}. Similarly, for periodic axis bending
with spatial period $\Lambda$, $\mathcal{E}_x(z)$ is a periodic
function of $z$, and the condition of boundness of $\Omega(z)$ is
given by
\begin{equation}
\int_0^{\Lambda}d \xi \exp[-i \phi(\xi)]=0,
\end{equation}
which is precisely the condition of 'dynamic localization'
previously investigated in Refs.\cite{Lenz03,Longhi05}.

\subsection{Two-dimensional array}
For the two-dimensional waveguide array model (38), the moment
equations turn out to decouple into two set of equations, similar
to Eqs.(42-46), separately acting onto the $x$ and $y$ directions.
The evolution equations for the beam center of mass $\langle x
\rangle$, $\langle y \rangle$ are then given by
\begin{eqnarray}
\langle x(z) \rangle & = & \langle x(0) \rangle+2 {\rm Im} \left\{
q_{0x} \Omega_x^*(z) \right\} \\
\langle y(z) \rangle & = & \langle y(0) \rangle+2 {\rm Im} \left\{
q_{0y} \Omega_y^*(z) \right\}
\end{eqnarray}
where we have set
\begin{eqnarray}
\Omega_{x,y}(z) & = & \int_0^z d \xi \Delta_{x,y} \exp[-i
\phi_{x,y}(\xi)] \\
 \phi_{x,y}(z) & = & \int_0^z d \xi \mathcal{E}_{x,y}(\xi)
\end{eqnarray}
and
\begin{eqnarray}
q_{0x} & = & \sum_{n,m} c^*_{n,m}(0)c_{n+1,m}(0) \\
q_{0y} & = & \sum_{n,m} c^*_{n,m}(0)c_{n,m+1}(0).
\end{eqnarray}
Similarly, the beam sizes $w_x$ and $w_y$, defined as
\begin{eqnarray}
w_x(z) & = & \sqrt{\langle x^2(z) \rangle-\langle x(z) \rangle^2 }
\\
w_y(z) & = & \sqrt{\langle y^2(z) \rangle-\langle y(z) \rangle^2
},
\end{eqnarray}
are calculated using Eqs.(58-59) and the following evolution
equations for $\langle x^2(z) \rangle$ and $\langle y^2(z)
\rangle$
\begin{equation}
\langle x^2(z) \rangle  =  \langle x^2(0) \rangle+2|\Omega_x|^2+2
{\rm Re} \left\{q_{1x}\Omega_x-q_{2x}\Omega_x^2 \right\}
\end{equation}
\begin{equation}
 \langle y^2(z) \rangle  =  \langle y^2(0)
\rangle+2|\Omega_y|^2+2 {\rm Re}
\left\{q_{1y}\Omega_y-q_{2y}\Omega_y^2 \right\}
\end{equation}
where we have set
\begin{eqnarray}
q_{1x} & = & i \sum_{n,m}
(2n-1)c^*_{n,m}(0)c_{n-1,m}(0) \\
q_{2x} & = & \sum_{n,m}c^*_{n,m}(0) c_{n-2,m}(0) \\
q_{1y} & = & i \sum_{n,m}
(2m-1)c^*_{n,m}(0)c_{n,m-1}(0) \\
q_{2y} & = & \sum_{n,m}c^*_{n,m}(0) c_{n,m-2}(0).
\end{eqnarray}

\section{Shape-invariant discrete beams}
 The existence of
shape-invariant beams, i.e. families of field distributions that
propagate without changing their functional shape, is well-known
for paraxial propagation in Gaussian optics or in continuous
lensguide media (see, for instance, \cite{Siegman}). Here we
address the related problem of investigating the existence of
shape-invariant discrete beams, i.e. field distributions that do
not change their functional shape when propagating along waveguide
arrays with arbitrarily curved optical axis. This is a rather
challenging problem because  no general method capable of
constructing shape-invariant beams seems to be available. However,
for the simple waveguide array models considered in the previous
section, a family of shape-invariant discrete beams can be
introduced in a simple manner. Owing to their functional form,
such beams are referred to as discrete Bessel beams.

\subsection {Discrete Bessel beams in one-dimensional arrays}
 Let us consider
a one-dimensional waveguide array with an arbitrarily curved
optical axis. In the tight-binding and nearest neighboring
coupling approximations, light propagation is described by the
following set of coupled-mode equations
\begin{equation}
i \dot{c}_n=-\Delta(c_{n+1}+c_{n-1})- n f(z) c_n
\end{equation}
where $f(z)$ describes the rate of transverse index gradient
induced by waveguide bending \cite{Longhi05}, lumped waveguide
tilting \cite{Eisenberg00} or locally imposed phase changes among
adjacent waveguides \cite{SzameitAPL08} as discussed previously.
Let us fist observe that, if $c_n(z)$ is a solution to Eq.(72)
corresponding to a given initial field distribution $c_n(0)$, then
for an arbitrary integer $n_0$
\begin{equation}
g_n(z)=c_{n-n_0}(z) \exp \left\{ i n_0 \int_0^z d \xi f(\xi)
\right\}
\end{equation}
is the solution to Eq.(72) corresponding to the translated initial
field distribution $g_n(0)=c_{n-n_0}(0)$. Therefore, apart from an
unimportant phase change, shape-invariant beams remain invariant
for an arbitrary transverse translation on the lattice.\\
Let us tentatively search for a solution to Eq.(72) of the form
\begin{equation}
c_n(z)=J_n(\alpha) \exp(-i \sigma n)
\end{equation}
where $J_n$ is the Bessel function of first kind of order $n$, and
$\alpha=\alpha(z)$, $\sigma=\sigma(z)$ are unknown functions which
depend on propagation distance $z$, but not on lattice site $n$.
Note that, as $\sum_n n|J_n(\alpha)|^2=0$ and $[\sum_n
|J_n(\alpha)|^2n^2]=\alpha^2/2$, the parameter $\alpha$ is related
to the beam size $w_x$ [Eq.(6)] by the simple relation
$w_x=\alpha/{\sqrt 2}$, whereas $\sigma$ defines a transverse tilt
of the beam 'phase front'. Substitution of Eq.(74) into Eq.(72)
and taking into account the identities of Bessel functions
$J_{n+1}(\alpha)+J_{n-1}(\alpha)=(2 n / \alpha) J_n(\alpha)$ and
$J_{n-1}(\alpha)-J_{n+1}(\alpha)=2J'_n(\alpha)$, one obtains that
Eq.(74) is indeed a solution to Eq.(72) provided that $\alpha$ and
$\sigma$ satisfy the coupled equations
\begin{eqnarray}
\dot{\alpha} & = & -2 \Delta \sin \sigma \\
\dot{\sigma} & = & -\frac{2 \Delta}{\alpha} \cos \sigma -f.
\end{eqnarray}
Owing to the functional form of $c_n$, we will refer such
shape-invariant beams to as discrete Bessel beams. Let us define a
complex-$q$ parameter for the discrete Bessel beam (74) according
to
\begin{equation}
q(z)=\alpha(z) \exp[i\sigma(z)]
\end{equation}
so that the modulus of the complex $q$ parameter gives the beam
spot size at propagation distance $z$, whereas its phase
corresponds to the phase front gradient. From Eqs.(75) and (76)
one readily obtains for the complex $q$ parameter the following
simple evolution equation
\begin{equation}
\frac{d q}{dz}=-2i\Delta-if(z) q.
\end{equation}
The general solution to Eq.(78), for a given initial value $q(0)$
at the $z=0$ input plane, is given by
\begin{equation}
q(z)=\exp[-i \phi(z)] \left\{q(0)-2i \int_0^z d \xi \Delta \exp[i
\phi(\xi)] \right\}
\end{equation}
where
\begin{equation}
\phi(z)=\int_0^z d \xi f(\xi).
\end{equation}
The propagation of a discrete Bessel beam along a curved waveguide
array is thus reduced to the propagation of its complex $q$
parameter,
 which plays an analogous role of
the complex-$q$ parameter for Gaussian beams in lensguide media.
The propagation law of the $q$ parameter admits of a simple
geometrical interpretation in the complex $q$ plane. According to
Eq.(78), for an infinitesimal propagation distance $\delta z$ the
change of $q(z)$ is given by the superposition of the two paths AB
and BC shown in Fig.2(a). The path AB, of length $2 \delta z
\Delta$, accounts for discrete diffraction and corresponds to a
change of $q(z)$  along the imaginary $q$ axis; the path BC is due
to the transverse index gradient which produces a clockwise
rotation by the angle $\delta \gamma=f(z) \delta z$ around the
origin O of the complex plane. It is interesting to note that,
since $J_n(0)=\delta_{n,0}$, for $q(0)=0$ the discrete Bessel beam
(74) reduces to the well-known impulse response of a tight-binding
array with nearest neighbor couplings (see, for instance, \cite{Dunlap86}).\\
To appreciate the usefulness of the $q$-parameter description and
some properties of discrete Bessel beams, let us now discuss a few examples
 and applications.\\
 \\
 \begin{figure}
\includegraphics[scale=0.9]{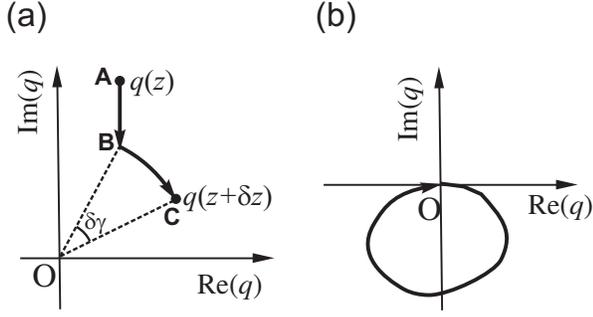}
\caption{(a) Geometric construction of the evolution of the
complex $q$ parameter for an infinitesimal propagation distance
$\delta z$. The length of the segment AB is $2\Delta \delta z$,
whereas the rotation angle is $\delta \gamma=f(z) \delta z$. The
points $A$ and $C$ correspond to $q(z)$ and $q(z+\delta z)$,
respectively. (b) Geometric representation of a self-imaging
array: the path followed by the complex parameter $q(z)$, starting
from the origin $O$, is closed.}
\end{figure}
{\it Propagation of discrete Bessel beams in homogeneous
arrays.}\\
For a homogeneous array ($f=0$), the propagation law of the
complex-$q$ parameter is simply given by
\begin{equation}
q(z)=q(0)-2i\Delta z.
\end{equation}
If we assume, for the sake of definiteness, that at the input
plane $z=0$ the phase front of the beam is flat, i.e.
$q(0)=\alpha(0)=\alpha_0$ real valued,  the following propagation
laws for beam size $\alpha$ and beam phase tilt $\sigma$ are
derived
\begin{eqnarray}
\alpha(z) & = & \alpha_0 \sqrt{1+\left( \frac{2 \Delta
z}{\alpha_0} \right)^2} \\
\sigma(z) & = & - {\rm arctan}\left( \frac{2 \Delta
z}{\alpha_0}\right).
\end{eqnarray}
From Eq.(82) we may introduce, as for Gaussian beams propagating
in free space \cite{Siegman}, the Rayleigh range $z_R$ and
divergence angle $\theta_d$ such that $\alpha(z_R)= \sqrt{2}
\alpha_0$ and $\theta_d= \lim_{z \rightarrow \infty} \alpha(z)/z$,
i.e.
\begin{eqnarray}
z_R & = & \frac{\alpha_0}{2 \Delta} \\
\theta_d & = & 2 \Delta.
\end{eqnarray}
It should be noted that, as opposed to the case of Gaussian beams
in free space -for which the Rayleigh range $z_R$ is proportional
to the {\it square} of the spot size $\alpha_0$ at the beam waist
and the diffraction angle $\theta_d$ is inversely proportional to
$\alpha_0$- for discrete Bessel beams the Rayleigh range $z_R$ is
proportional to the spot size $\alpha_0$ at the beam waist whereas
the divergence angle is {\it independent} of the beam spot size
and always equal to the  diffraction cone angle introduced in
Sec.II.D. This peculiar property is closely related to the very
general result, proven in Sec.II.D, that the far field of discrete
diffraction in a homogenous waveguide array is peaked at the
observation angles corresponding to the
flattest points (self-collimation points) of the band dispersion curve.\\
\\
 \begin{figure}
\includegraphics[scale=0.75]{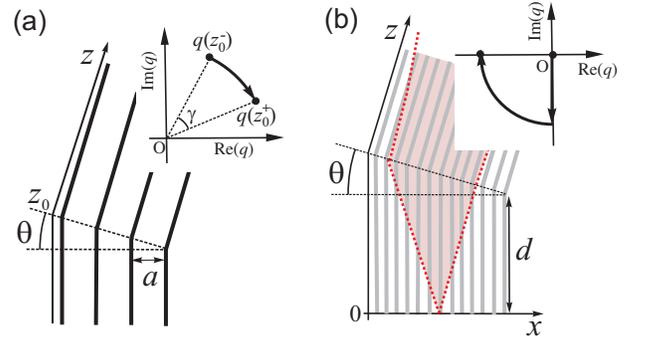}
\caption{(color online) (a) Schematic of a one-dimensional
waveguide array with a tilt of waveguide axis at $z=z_0$, and the
transformation induced on the complex $q$ parameter by the tilt
(inset). (b) Principle of beam collimation via waveguide axis tilt
[tilt angle $\theta=\lambda/(4an_s)$], and path followed by the
complex $q$ parameter for single waveguide input excitation from
$z=0$ to $z=d^+$ (inset).}
\end{figure}

{\it Transformation of a discrete Bessel beam through a waveguide
axis tilt.} A tilt of the waveguide axis at $z=z_0$ by a (small)
angle $\theta$ corresponds to impressing a phase shift
\begin{equation}
\gamma=\frac{2 \pi} {\lambda} a \theta n_s
\end{equation}
 between adjacent
waveguides, where $a$ is the waveguide spacing and $n_s$ the
effective index of propagating modes [see Fig.3(a)]. Light
propagation across the tilt can be thus modelled by assuming
$f(z)= \gamma \delta(z-z_0)$ in Eq.(72), and its effect on the
complex $q$ parameter is to produce a rotation around the origin
of the complex plane by an angle $\gamma$ [see the inset of
Fig.3(a)].\\
A tilt of the waveguide axis may be used to 'collimate' a discrete
beam, as schematically shown in Fig.3(b). Here a single waveguide
is initially excited at the input plane, and after a propagation
distance $d$ the axis of the array is tilted by an angle
$\theta=\lambda/(4an_s)$ such that $\gamma=\pi/2$. The 90$^o$
rotation of the $q$ parameter in the complex plane due to axis
bending [see the inset in Fig.3(b)] brings the $q$ parameter on
the real axis, with a zero phase gradient $\sigma=0$ and an
enlarged  beam size $\alpha=2 \Delta d$. The axis tilt thus plays
a similar role of a collimating lens for a diverging Gaussian
beam. Note however that, contrary to a conventional lens, the
tilting angle $\theta$ to achieve beam collimation is independent
of the distance $d$ between source point (at $z=0$) and the lens
plane ($z=d$). Figure 4 shows an example of beam collimation in a
6-cm-long one-dimensional array as obtained by  numerical analysis
of the scalar wave equation for the electric field envelope
$E(x,z)$ propagating in the structure based on a standard beam
propagation method. Figure 4(a) shows a pseudocolor map of the
intensity beam evolution $|E(x,z)|^2$ along the structure when a
single waveguide is excited in its fundamental mode at the input
plane $z=0$ and the waveguide axis is tilted at a distance $d=2$
cm from the input plane [horizontal dotted curve in Fig.4(a)]. The
refractive index profile $n(x)$ used in the simulations is
depicted in Fig.4(b), and the values of other parameters are
$\lambda=1.55 \; \mu$m, $n_s=1.52$, and $a=11 \; \mu$m,
corresponding to a tilting angle $\theta=\lambda/(4a n_s) \simeq
23.2 \; {\rm mrad}$. For the sake of readability, the intensity
distribution is plotted with the waveguide axis $z$ unfolded along
a straight line. Note that the numerical results provide a
realistic behavior of beam propagation beyond the couple-mode
equation approximation, accounting for radiation losses and
coupling to higher-order bands due to axis bending. These latter
effects,
 however, are very small for the parameter values adopted in the simulations, and the coupled-mode
 equation model works fine.\\
\\
 \begin{figure}
\includegraphics[scale=1.1]{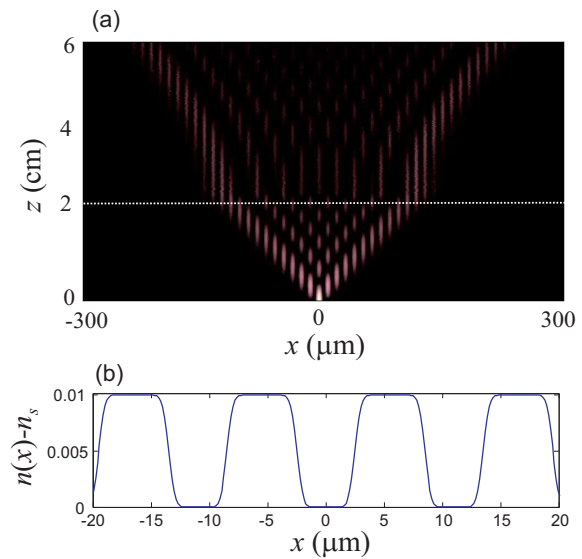}
\caption{(color online) (a) Pseudocolor map of beam intensity
propagation in a waveguide array with one axis tilting as obtained
by numerical simulations, showing beam collimation. (b) Refractive
index profile of the waveguide array used in the numerical
simulations. The values of other parameters are given in the
text.}
\end{figure}
 {\it A geometric interpretation of the self-imaging condition and polygonal Bloch oscillations}.
 An array of length $d$ shows a self-imaging property (also
referred to as diffraction cancellation or dynamic localization) ,
whenever $|c_n(d)|^2=|c_n(0)|^2$ for any initial field
distribution. The dynamic localization condition has a rather
simple geometric interpretation in the complex $q$ plane. In fact,
if the array is excited in waveguide $n=0$, $q(0)=0$ and to
achieve self-imaging after a propagation distance $d$ one has
necessarily to have $q(d)=q(0)=0$, i.e the path described by the
complex $q$ parameter, starting from the origin O of the complex
plane, should be closed [see Fig.2(b)]. Owing to the translational
invariance of discrete Bessel beams [Eq.(73)], this condition is
also sufficient. From Eq.(79), the closed-path condition
$q(d)=q(0)=0$ yields
\begin{equation}
\int_0^d dz \exp[i \phi(z)]=0
\end{equation}
which is precisely the condition for dynamic localization derived
originally by Dunlap and Kenkre in Ref.\cite{Dunlap86}.\\
An application of the geometric condition of dynamic localization
is that of polygonal Bloch oscillations. Let us consider a
waveguide array whose axis forms an (open) polygonal curve of
large (mean) radius $R$ made of a sequence of straight segments of
same length $b$ and with tilt angle $\theta$, as shown in
Fig.5(a). The function $\phi(z)$, defined by Eq.(80), is thus a
staircase function, which increases in steps of $\gamma=(2
\pi/\lambda)a \theta n_s$ [see Eq.(86)] at $z=b, 2b,3b,...$ (the
coordinate $z$ is measured along the polygonal curve). After a
propagation $d=(N+1)b$ from the input $z=0$ plane, where $N$ is an
integer number, it then follows that
\begin{equation}
\int_0^d dz \exp[i \phi(z)]=b \sum_{n=0}^N \exp(i \gamma n).
\end{equation}
The sum of complex numbers (phasors) on the right hand side of
Eq.(88) can be done analytically and has a well-known geometric
interpretation; in particular, if $\gamma$ satisfies the condition
$\gamma= 2 \pi/(N+1)$, i.e. if the tilt angle $\theta$ is given by
\begin{equation}
\theta=\frac{\lambda}{a n_s (N+1)}
\end{equation}
the sum on the right hand side of Eq.(88) vanishes, and the
condition for self-imaging is attained. An example of the
self-imaging property of a polygonal waveguide array is shown in
Fig.5(b) for the case $N=5$. The figure depicts a characteristic
breathing mode corresponding to a single waveguide excitation at
the input plane. The waveguide array parameters are the same as in
Fig.4, and a sequence of axis tilts are placed at distances $b=1$
cm one to the next. The tilt angle $\theta$, chosen according to
Eq.(89), is $\theta \simeq 15.5 \; {\rm mrad}$, yielding a
self-imaging plane at $d=(N+1)b=6 \; {\rm cm}$, as clearly shown
in Fig.5(b). Note that, in the limit $b \rightarrow 0$, $N
\rightarrow \infty$ and $b / \theta \rightarrow R$ finite, the
polygonal of Fig.5(a) approximates an arc of a circumference of
radius $R$, and the condition (89) for self-imaging is satisfied
for a propagation distance
\begin{equation}
d=(N+1)b \rightarrow \frac{\lambda R}{n_s a}
\end{equation}
which is the spatial period of Bloch oscillations on a curved
waveguide array (radius of curvature $R$) previously considered in
Refs.\cite{Lenz99,Usievich04}. The usual Bloch oscillations on a
curved waveguide array may be therefore viewed as a limiting case
of Bloch oscillations on a polygonal array.
\begin{figure}
\includegraphics[scale=0.6]{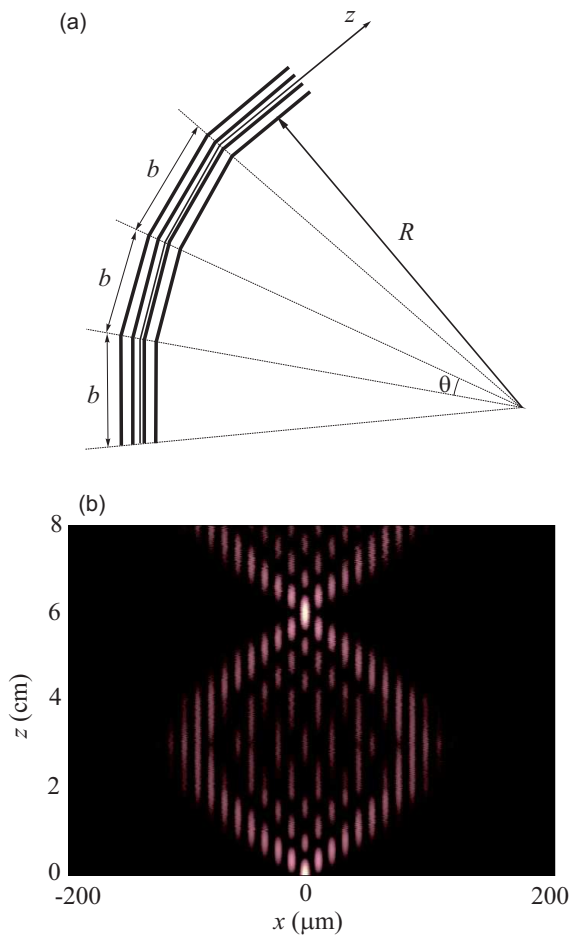}
\caption{(color online) (a) Schematic of a polygonal waveguide
array for the observation of Bloch oscillations.  (b) Pseudocolor
image of beam intensity propagation in a 8-cm-long polygonal array
showing a Bloch oscillation breathing mode. The refractive index
profile of the waveguide array used in the numerical simulations
is the same as in Fig.4(b). The values of other parameters are
given in the text.}
\end{figure}

\subsection {Discrete Bessel beams in two-dimensional arrays}
A simple extension of the analysis of Sec.IV.A can be done for a
two-dimensional rectangular-lattice waveguide array with
nearest-neighboring coupling when the diagonal coupling is
neglected. This model is described by the coupled mode equations
\begin{eqnarray}
i \dot{c}_{n,m}& = &
-\Delta_x(c_{n+1,m}+c_{n-1,m})-\Delta_y(c_{n,m+1}+c_{n,m-1})
\nonumber \\
& - & f_x(z) n c_{n,m}-f_y(z) m c_{n,m}
\end{eqnarray}
where $f_{x,y}(z)$ describe the rates of transverse index
gradients induced by waveguide bending or lumped waveguide axis
tilting along the $x$ and $y$ directions. Since Eqs. (91) admit of
separable solutions $c_{n,m}(z)=c_n(z) c_{m}(z)$, with $c_{n}(z)$
and $c_m(z)$ solutions to the one-dimensional problem (72) with
$\Delta=\Delta_{x,y}$ and $f(z)=f_{x,y}(z)$, a two-dimensional
discrete Bessel beam has the form
\begin{equation}
c_{n,m}(z)=J_n(\alpha_x) J_m(\alpha_y) \exp(-i\sigma_x n-\sigma_y
m).
\end{equation}
The complex-$q$ parameters of the beam along the $x$ and $y$
directions are defined by
\begin{equation}
q_x(z)= \alpha_x(z) \exp[i \sigma_x(z)] \; , \; q_y(z)=
\alpha_y(z) \exp[i \sigma_y(z)]
\end{equation}
and their evolution is ruled out by the equations
\begin{equation}
\dot{q}_{x,y}=-2i\Delta_{x,y}-if_{x,y}(z)
\end{equation}
which have a similar geometric interpretation as that discussed in
Sec.IV.A. The propagation properties of two-dimensional discrete
Bessel beams in homogeneous arrays, across tilted axis regions or
polygonal curves are the same as those investigated for
one-dimensional beams, and are therefore not further discussed
here.

\section{Conclusions}
In this work, a comprehensive study of discrete diffraction and
linear propagation of light in
 homogeneous and curved waveguide arrays has been presented.
 In particular, general laws describing beam spreading,
 beam decay and discrete far-field patterns in homogeneous arrays have been derived using the method of
 moments and the steepest descend method, and some remarks on the well-known
 self-collimation regime have been pointed out. In curved arrays and within the nearest neighboring
 coupling approximation, the  method of moments has been extended to describe the evolution of global
 beam parameters. This method provides an alternative means to algebraic operator techniques
 recently proposed in other physical contexts to  study general properties of Bloch oscillations \cite{B01,B02,B03}.
  Finally, a family of shape-invariant discrete beams -referred to as discrete Bessel beams owing to their functional
 form- has been introduced. It has been shown that propagation of such beams in curved waveguide arrays is simply
described by the evolution of a complex $q$ parameter, which plays
a similar role to the complex $q$ parameter used for Gaussian
beams in continuous lensguide media. A few applications of the $q$
parameter formalism are discussed, including beam collimation via
waveguide axis tilting, a geometric interpretation of the
self-imaging effect in waveguide arrays, and optical Bloch
oscillations on a polygonal array.


\begin{thebibliography}{}
\bibitem{notaS}
Analytical results can be also provided for a more general
two-dimensional lattice model, in which nearest diagonal coupling
terms are accounted for in the Hamiltonian (38) (see
Ref.\cite{B03}).  However, the resulting equations are very
cumbersome, and this case will not be considered here.
\end{thebibliography}
\end{document}